\documentclass{IEEEtran4PSCC}
\usepackage{nomencl}
\makenomenclature
\usepackage{siunitx}

\usepackage[caption=false,font=footnotesize]{subfig}	%	The option 'caption=false' is required to keep IEEE-style captions.
\usepackage{xcolor}

\definecolor{blue}{RGB}{31, 119, 180}    % Python plot blue
\definecolor{orange}{RGB}{255, 127, 14}  % Python plot orange
\definecolor{green}{RGB}{44, 160, 44}    % Python plot green
\definecolor{red}{RGB}{214, 39, 40}      % Python plot red
\definecolor{purple}{RGB}{148, 103, 189} % Python plot purple

\usepackage{threeparttable}
\usepackage{booktabs}

\usepackage{algorithm}
\usepackage{algorithmicx}
\usepackage{algpseudocode}
\usepackage{courier}

\usepackage{amsmath,amsfonts,amssymb,amsthm}
\usepackage{mathtools}
\usepackage{mathrsfs}	%	mathscr
\usepackage{subdepth}	%	uniform subscript position

\usepackage{array}	%	better appearance for array and tabular environments
\usepackage[inline]{enumitem} % enumerate stuff in line

\usepackage{hyperref}
\usepackage[sort,compress,capitalise]{cleveref}	%	...
\usepackage{url}

\usepackage{xcolor} % Coloured text
\usepackage{xspace} % Intelligent spacing (acronym commands) 
\usepackage{lipsum}

\usepackage{tikz}
\usetikzlibrary{babel}  % Avoid problems due to babel library
\usepackage[straightvoltages,nooldvoltagedirection]{circuitikz}
\usetikzlibrary{shapes,arrows}
\usepackage{verbatim}
\usepackage{multirow}

 \usepackage{booktabs}

\DeclareSIUnit{\rpm}{rpm}
\DeclareSIUnit{\var}{VAr}

{
\theoremstyle{plain}

}

{
\theoremstyle{definition}

}

\crefname{Hypothesis}{Hyp.}{Hyps.}
\Crefname{Hypothesis}{Hyp.}{Hyps.}

\crefname{Lemma}{Lemma}{Lemmata}
\Crefname{Lemma}{Lemma}{Lemmata}

\crefname{Definition}{Def.}{Defs.}
\Crefname{Definition}{Def.}{Defs.}

\newcommand{\DFT}[1][]{DFT\xspace}  % Discrete Fourier Transform (DFT)

\usepackage{bm}
\usepackage{mathtools}

\usepackage{xcolor}

\definecolor{blue}{RGB}{31, 119, 180}    % Python plot blue
\definecolor{orange}{RGB}{255, 127, 14}  % Python plot orange
\definecolor{green}{RGB}{44, 160, 44}    % Python plot green
\definecolor{red}{RGB}{214, 39, 40}      % Python plot red
\definecolor{purple}{RGB}{148, 103, 189} % Python plot purple % KETTNER

\ifCLASSINFOpdf
\else
   \usepackage[dvips]{graphicx}
\fi
\makeatletter
\let\old@ps@headings\ps@headings
\let\old@ps@IEEEtitlepagestyle\ps@IEEEtitlepagestyle
\def\psccfooter#1{%
    \def\ps@headings{%
        \old@ps@headings%
        \def\@oddfoot{\strut\hfill#1\hfill\strut}%
        \def\@evenfoot{\strut\hfill#1\hfill\strut}%
    }%
    \def\ps@IEEEtitlepagestyle{%
        \old@ps@IEEEtitlepagestyle%
        \def\@oddfoot{\strut\hfill#1\hfill\strut}%
        \def\@evenfoot{\strut\hfill#1\hfill\strut}%
    }%
    \ps@headings%
}
\makeatother

\psccfooter{%
        \parbox{\textwidth}{\hrulefill \\ \small{23rd Power Systems Computation Conference} \hfill \begin{minipage}{0.2\textwidth}\centering \vspace*{4pt} \includegraphics[scale=0.06]{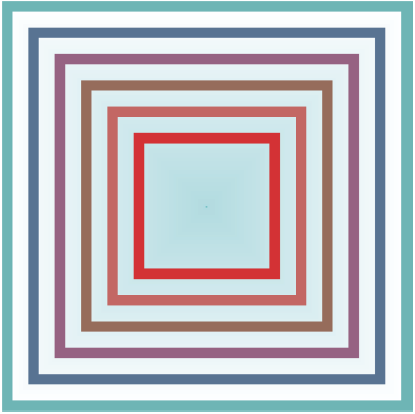}\\\small{PSCC 2024} \end{minipage} \hfill \small{Paris, France --- June 4 -- 7, 2024}}%
}

\begin{document}
\title{Experimental Investigation of Repurposed Kaplan Turbines as Variable-Speed Propellers for Maximizing Frequency Containment Reserve}

%% To specify the authors when (number of affiliations <= 2)
% \author{
% \IEEEauthorblockN{Francesco Gerini, Rachid Cherkaoui, Mario Paolone}
% \IEEEauthorblockA{Distributed Electrical System Laboratory (DESL) \\
% EPFL\\
% Lausanne, Switzerland\\
% \{francesco.gerini, rachid.cherkaoui, mario.palone\}@epfl.ch}
% \and
% \IEEEauthorblockN{Martin Seydoux, Elena Vagnoni}
% \IEEEauthorblockA{Technology Platform for Hydraulic Machines T(PTMH) \\
% EPFL\\
% Lausanne, Switzerland\\
% \{martin.seydoux, elena.vagnoni\}@epfl.ch}
% }

% To specify the authors when (number of affiliations > 2)
\author{\IEEEauthorblockN{Francesco Gerini\IEEEauthorrefmark{2},
Elena Vagnoni\IEEEauthorrefmark{1}, 
Martin Seydoux\IEEEauthorrefmark{1},
Rachid Cherkaoui\IEEEauthorrefmark{2} and
Mario Paolone\IEEEauthorrefmark{2}}
\IEEEauthorblockA{\IEEEauthorrefmark{2} Distributed Electrical System Laboratory (DESL) \\
EPFL,
Lausanne, Switzerland\\ }% \{francesco.gerini, rachid.cherkaoui, mario.palone\}@epfl.ch}
\IEEEauthorblockA{\IEEEauthorrefmark{1} Technology Platform for Hydraulic Machines (PTMH)\\
EPFL,
Lausanne, Switzerland\\ }%\{martin.seydoux, elena.vagnoni\}@epfl.ch}
}

\maketitle

\begin{abstract}
This study explores the practical viability of repurposing aging Kaplan turbines into variable-speed propellers by employing full-size frequency converters. The motivation behind this approach is to improve the provision of \emph{Frequency Containment Reserve} (FCR) while reducing fatigue in the Kaplan blades servomechanism. We evaluate the performance of these modified Kaplan turbines against the one of another hydro asset composed of the same Kaplan turbine hybridized with a \emph{Battery Energy Storage System} (BESS). Experiments are conducted on a one-of-its-kind reduced-scale model testing platform. Our findings reveal that Kaplan turbines repurposed as variable-speed propellers exhibit similar dynamic response characteristics compared to the standalone Kaplan operation, with the added benefit of effectively eliminating blade movements. Furthermore, the ability to control the speed increases the hydraulic efficiency for certain operating points. In summary, investment in variable speed technology emerges as a viable alternative to BESS-based hydropower hybridization.
\end{abstract}

\begin{IEEEkeywords}
Variable Speed, Battery Energy Storage System, Frequency Containment Reserve, Hydropower, Run-of-River Power Plant
\end{IEEEkeywords}

% Use this to place sponsorships
\thanksto{\noindent This work is funded by the XFLEX HYDRO project. The XFLEX HYDRO project has received funding from the European Union’s Horizon 2020 research and innovation programme under grant agreement No. 857832. }
\section{Introduction}
% Introduction
The gradual integration of renewable energy sources into Europe's power grid is crucial for reducing carbon emissions. Among these sources, hydropower plays a significant role, substantially contributing to the electricity generation mix \cite{entsoe_statistical_2022}. Beyond its clean energy production, hydropower offers essential support for grid balancing and flexibility, providing \emph{ Frequency Containment Reserve} (FCR) \cite{european_commission_commission_2017}. Continuous adjustments, required to meet the FCR dynamics, can stress hydropower assets \cite{yang_burden_2018}. This stress is particularly noticeable in double-regulated hydraulic machines such as Kaplan turbines, where servomotors control the positioning of both guide vanes and blades for efficient power production through a variety of heads and discharges.

% Literature Review
To address these challenges, researchers have explored hybrid solutions that combine \emph{Battery Energy Storage Systems} (BESS) with hydroelectric plants to provide rapid and dynamic responses \cite{kadam_hybridization_2023,valentin_hybridization_2022,cassano_model_2022,gerini_enhanced_2023}. In particular, \cite{gerini_enhanced_2023} has shown that BESS hybridization is effective in mitigating aging of servomotor mechanisms and improving the quality of FCR provision by the power plant. However, it comes at a notable initial \emph{Capital Expenditure} (CAPEX) cost. In Europe, notable examples of BESS hybridization include an 8 MW/10 MWh BESS in Austria \cite{nidec-industrial_8_2020}, paired with an 8 MW turbine, and a 650 kW/350 kWh BESS integrated with a 35 MW Kaplan turbine at the Vogelgrun (FR) power plant \cite{valentin_hybridization_2022}. Although the latter investment is relatively modest in size, the former requires a BESS with rated power equal to the one of the \emph{Hydropower Plant} (HPP), resulting in a considerably higher cost. This substantial investment is justified by the need to maintain the operational integrity of a power plant that would otherwise require extensive refurbishment.

% Our approach
This study discusses an alternative approach by investigating the retrofit of existing Kaplan turbines with full-size frequency converters, enabling \emph{Variable-Speed} (VARspeed) control \cite{fraile-ardanuy_variable-speed_2006} while fixing the blade angle. This modification consists of repurposing the Kaplan turbine as a variable-speed propeller. The utilization of a full-size frequency converter becomes particularly compelling in RoR HPPs equipped with Kaplan turbines, due to their typical rated powers in the tens of megawatts range. Unlike BESS hybridization, VARspeed operation requires only CAPEX investments associated with power electronics, with no requirement for energy storage. The concept of VARspeed operation for hydraulic turbines has been extensively examined in the academic literature, particularly focusing on Francis turbines and reversible pump-turbines \cite{mercier_provision_2017,yang_advantage_2019,chazarra_optimal_2018}. Moreover, numerous studies have provided compelling evidence for improved FCR provision through VARspeed technology \cite{reigstad_optimized_2020,reigstad_nonlinear_2021,dreyer_pushing_2022}. These studies have also addressed control optimization challenges associated with harnessing the full potential of VARspeed technology.

In this study, we conduct a comprehensive \textcolor{black}{technical} comparison between the BESS hybridization of a Kaplan turbine and the VARspeed control of the same unit used as a propeller, i.e., with fixed blades. Both VARspeed control and BESS hybridization techniques undergo extensive experimentation using an advanced testing platform equipped with a grid emulator and a full-size frequency converter connected to a synchronous electrical machine \cite{gerini_experimental_2021}. The grid emulator recreates various grid scenarios, enabling an assessment of the VARspeed system's FCR response and its comparison with the BESS hybrid HPP. During the test, a dedicated set of hydraulic measurements allows for estimating hydraulic and global efficiency, as well as measuring stresses on mechanical components of the hydraulic machine.

The paper is organized as follows. Section \ref{sec:Problem Statement} presents the problem within its context. Section \ref{sec:Design of Experiments} presents a detailed description of the design of experiments. Section \ref{sec:Experimental Validation} provides and discusses the experimental results. Finally, Section \ref{sec:Conclusion} summarizes the original contributions and main results of the article and proposes perspectives for further research activities.

\section{Problem Statement}
\label{sec:Problem Statement}
Let us consider a \emph{Run-of-River} (RoR) HPP equipped with a Kaplan turbine. The HPP operates under a daily dispatch plan $P^{\text{disp}}$ and is obliged to provide FCR with a fixed droop\footnote{Many European Transmission System Operators (TSOs) procure their FCR from a jointly operated market that allows for 4-hour adjustments, as indicated by \cite{entsoe_tsos_2018}. However, it should be noted that the settings for HPP droop controls often remain unchanged for extended periods, spanning years or even decades, due to contractual agreements. As a result, these settings are considered constant variables in the scope of this research.} characteristic $\sigma_f$. As the unit approaches the end of its operational life, the HPP owner aims to mitigate the aging of the servomechanism while maintaining a continuous provision of FCR. In this context, two distinct approaches are considered: BESS hybridization of the Kaplan unit, and VARspeed control on the same unit but retrofitted as a propeller, i.e. with fixed blades. The primary objective of this study is to conduct a technical assessment of these two approaches, with a specific focus on the following objectives:
\begin{enumerate}[label=\roman*)]
    \item Ensuring dispatch tracking and high-quality FCR provision with a rapid response time, in strict compliance with stringent FCR requirements \cite{entsoe_all_2019,swissgrid_ltd_conditions_2009};
    
    \item Reducing the number of movements and the mileage of the hydropower servomechanisms, with particular attention to the blade servomechanism;
         
    \item Maximising the overall efficiency of the system.
\end{enumerate}

\begin{figure}[t]
  \centering
  \subfloat[Schematics of the PTMH PF3 power grid used in BESS hybridization mode]{%
    \includegraphics[width=\linewidth]{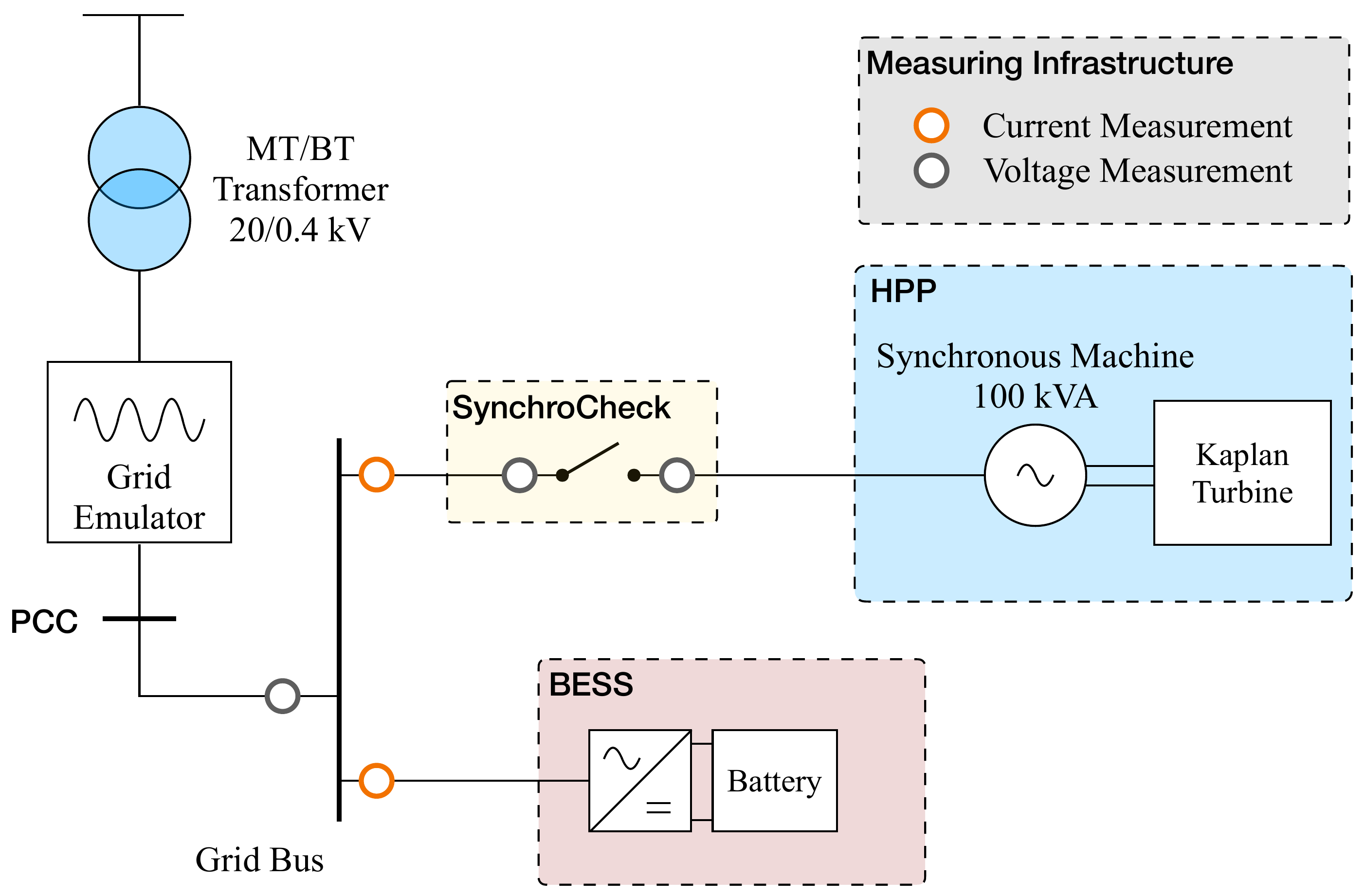}
    \label{fig:Schematics_BESS}%
  }\\
  \subfloat[Schematics of the PTMH PF3 power grid used in VARspeed mode]{%
    \includegraphics[width=\linewidth]{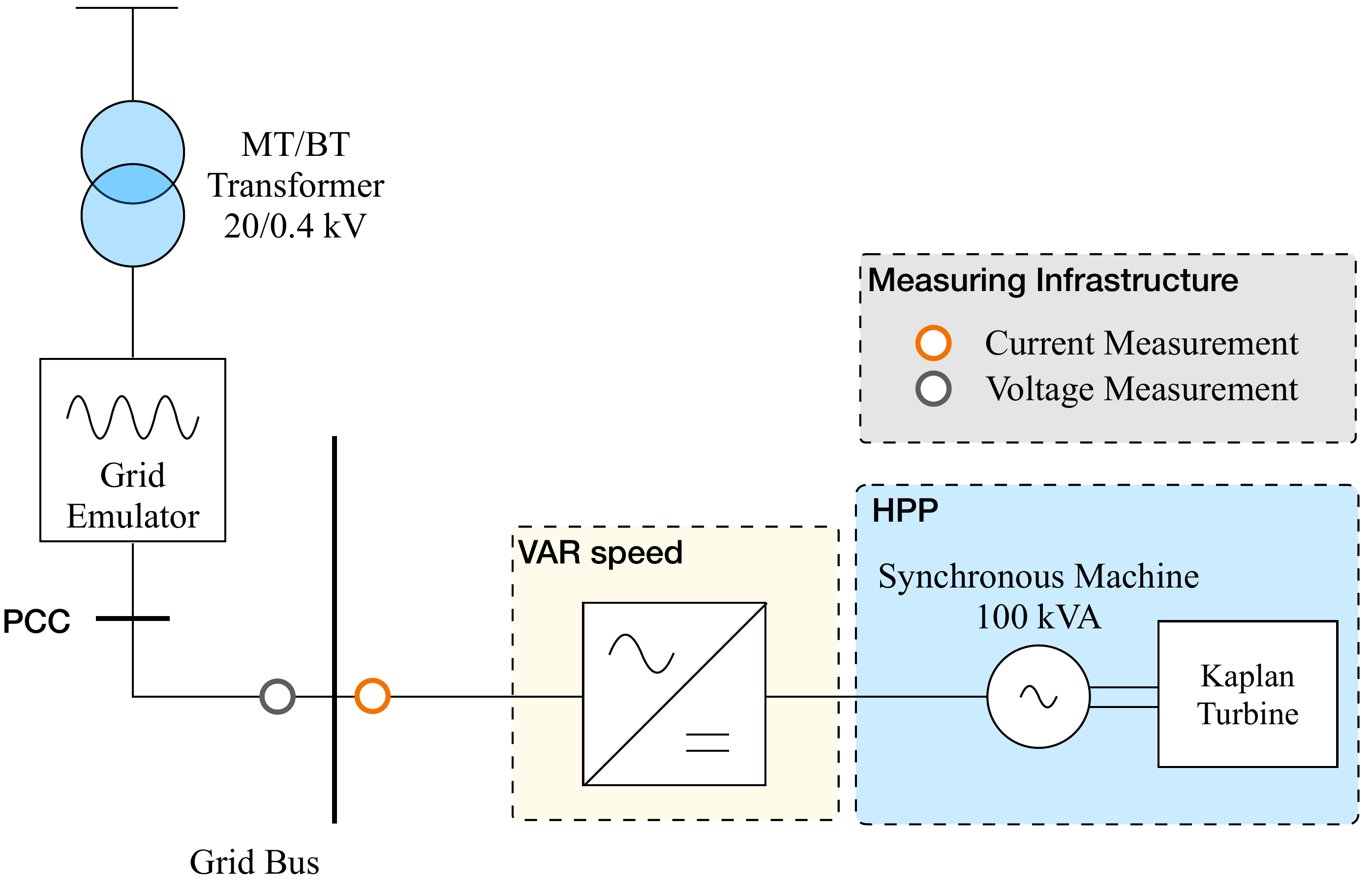}
    \label{fig:Schematics_VarSpeed}%
  }
  \caption{Schematics of the PTMH PF3 power grid in different operating modes}
  \label{fig:Schematics_Comparison}
\end{figure}

The configuration of the two systems is illustrated in \cref{fig:Schematics_Comparison}. The following section provides details regarding the experimental assessment of the two approaches.

\section{Design of Experiments}
\label{sec:Design of Experiments}
\subsection{Experimental Setup}
\label{sec:Experimental_setup}
\begin{figure}
\centering
\includegraphics[width=\linewidth]{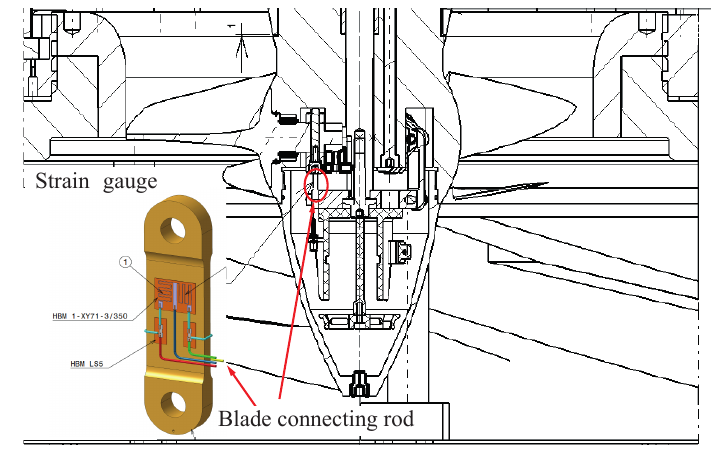}
\caption{HBM strain gauge on the connecting rod of one blade of the Kaplan Turbine}
\label{fig:StrainGauges}
\end{figure}

The experimental platform used for this study modernizes the existing testing platform in the \emph{Plateforme Technologique Machines Hydrauliques} (PTMH), at EPFL. The upgraded platform is a closed-loop test rig allowing the performance assessment of hydraulic machines with 0.2 \% accuracy, complying with the IEC60193 standard for the testing of reduced-scale physical models \cite{standard_iec_and_others_iec_2019}. Two centrifugal pumps allow for generating a maximum head of \SI{100}{\meter} and a maximum discharge of \SI{1.4}{\cubic\meter\per\second}. The hydroelectric unit governing systems is built as a standard speed governor \cite{grigsby_power_2007} with a frequency droop $\sigma_f =  \SI{125}{\kilo\watt\per\hertz}$ and a $\SI{2}{\milli\hertz}$ dead band. The choice of droop is dictated by the need for similarity with the prototype of the reduced-scale model turbine, installed in Vogelgrun (FR) \cite{kadam_hybridization_2023}. The electrical measurement infrastructure employs a distributed sensing system based on \emph{Phasor Measurement Units} (PMUs) relying on the \emph{enhanced-Interpolated DFT} (e-IpDFT) algorithm developed in \cite{romano_enhanced_2014}. This system allows real-time acquisition of accurate power flow data thanks to the high reporting rate of the PMUs (\SI{50}{\hertz}) and remarkable precision, with a standard deviation equivalent to $0.001$ degrees (approximately 18 $\mu$rad) and error in frequency estimation $\le \SI{0.4}{\milli\hertz}$. A set of switches is controlled to modify the grid topology \cite{gerini_experimental_2021} and test both BESS hybridization mode and VARspeed mode, as shown in \cref{fig:Schematics_BESS} and \cref{fig:Schematics_VarSpeed}, respectively.
Hydraulic measurements are performed on a 1:20 reduced-scale physical model of a Kaplan turbine with a specific speed $v=1.53$ defined as follows:
\begin{equation}
    v=\omega \cdot \frac{Q^{1 / 2}}{\pi^{1 / 2} \cdot(2 \cdot E)^{3 / 4}}
\end{equation}
where $Q$ is the hydraulic discharge, $E$ the machine specific energy and $\omega$ the angular speed in \SI{}{\radian\per\second}. The reduced-scale model includes the upstream inlet pipe, the spiral case, the 4-blade runner with an external diameter $D=\SI{0.34}{\meter}$, and the draft tube consisting of the cone, the elbow, and the diffuser. An electric actuator is used to move the blades of the Kaplan turbine and define the operating point. 
The turbine, with a rated power of 50 kW, is connected to the grid bus with a 4-pole synchronous machine, rated 100 kVA. The frequency of the grid is imposed by a grid emulator with a nominal power rating of 100 kVA. As the turbine is connected to the grid through a synchronous machine, a synchro-check mechanism is necessary for the BESS hybridization tests. The comprehensive list of measurements is detailed in Table \ref{tab:measurements}. Particular attention is paid to the measurement of the stresses on the connecting rod of one blade. The measure is performed with a \emph{HBM strain gauge} to evaluate the torque on the blade connecting rod and the related wear and tear. The strain gauge setup is visible in \cref{fig:StrainGauges}. As outlined in \cite{presas_use_2021}, strain gauge testing emerges as the only reliable approach to acquire static and dynamic stress information from the turbine runner. Over recent years, this method has gained widespread acceptance as a standard practice for newly installed runners, as highlighted in \cite{presas_fatigue_2019}.
All hydraulic measurements are performed at a sampling rate of 1 Hz for the full duration of the 12-hour testing scenarios.
\begin{table*}[ht]
    \centering
    \caption{Measurements on the PTMH-PF3 Testing Platform}
    \begin{tabular}{l c l c l c}
    \toprule
    \multicolumn{2}{c}{\textbf{Hydropower}} & \multicolumn{2}{c}{\textbf{BESS}} & \multicolumn{2}{c}{\textbf{PMU}} \\
    \cmidrule(lr){1-2} \cmidrule(lr){3-4} \cmidrule(lr){5-6}
    \textbf{Name} & \textbf{Measurement Unit} & \textbf{Name} & \textbf{Measurement Unit} & \textbf{Name} & \textbf{Measurement Unit} \\
    \midrule
    Discharge & \SI{}{\cubic\meter\per\second} & State of Charge & \SI{}{\percent} & Voltage Magnitude & \SI{}{\volt}\\
    Turbine Speed & \SI{}{\per\min} & DC Current & \SI{}{\ampere} & Voltage Angle & \SI{}{\radian} \\
    Head & \SI{}{\meter} & DC Power & \SI{}{\watt} & Current Magnitude & \SI{}{\ampere} \\
    Hydraulic Efficiency & \SI{}{\percent} & DC Voltage & \SI{}{\volt} & Current Angle & \SI{}{\radian} \\
    Shaft Torque & \SI{}{\newton\meter} & Active power & \SI{}{\watt} & Active Power & \SI{}{\watt} \\
    Guide Vanes Opening (GVO) & \SI{}{\deg} & Reactive Power & \SI{}{\var} & Reactive Power & \SI{}{\var} \\
    Runner Blade Angle (RBA) & \SI{}{\deg} & Temperature & \SI{}{\degreeCelsius} & Frequency & \SI{}{\hertz} \\
    Runner Blade Torque (RBT) & \SI{}{\newton\meter} & & & & \\
    \bottomrule
    \end{tabular}
    \label{tab:measurements}
\end{table*}
\subsection{Experimental Test}
\label{sec:Experimental_tests}
\begin{figure*}
\centering
\includegraphics[width=\linewidth]{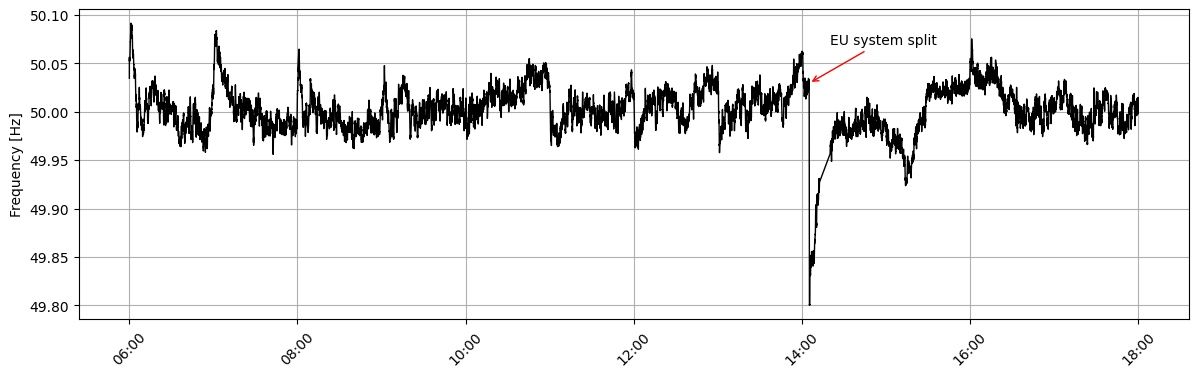}
\caption{Frequency Time Series for the test. Measurement from SwissGrid during the system split of the 8th of January 2021 \cite{ics_investigation_expert_panel_continental_2021}}
\label{fig:frequency}
\end{figure*}

A series of 12-hour tests are performed under equal conditions, with frequency time series enforced at the \emph{Point of Common Coupling} (PCC) by the grid emulator. The frequency data corresponds to measurements taken on January 8th, 2021, when the ENTSO continental Europe synchronous area experienced a system split, and it is illustrated in \cref{fig:frequency}. This particular time frame is chosen to ensure the inclusion of typical daily frequency patterns, for the first \SI{8}{\hour} of the test, and also to subject the system to more challenging scenarios in the remaining 4 hours. More information on the system split event can be found in \cite{ics_investigation_expert_panel_continental_2021}.
Under the described grid conditions, the following tests are conducted:
\begin{enumerate}[label=\alph*)]
    \item Kaplan unit operating alone ('\emph{Only Hydro}').
    \item Hybrid Kaplan unit with a \SI{5}{\kilo\watt}/\SI{5}{\kilo\watt\hour} BESS.
    \item Hybrid Kaplan unit with a \SI{9}{\kilo\watt}/\SI{9}{\kilo\watt\hour} BESS.
    \item VARspeed Propeller
\end{enumerate} where the BESS-hybridized HPP are controlled with the \emph{Double Layer MPC} (DLMPC) proposed in \cite{gerini_enhanced_2023}. All the tests are performed considering a flat dispatch plan of \SI{27}{\kilo\watt}, constant for the 12 hours, to ensure that the wear and tear analysis of the movements is only related to FCR support. Moreover, the head of the system is kept constant at $\SI{10}{\meter}$. The governing systems for the Kaplan control and the VARspeed propeller control are presented in \cref{sec:CAMcurve} and \cref{sec:VARspeedGovernor}, respectively.

\subsection{Key performance Indicators}
\label{sec:KPI}
\subsubsection{FCR provision quality}
To assess the quality of the FCR provision, together with the dispatch tracking ability, the active power flow $P^{\text{pcc}}$ measured at the PCC is considered. For every time step $k$ the expected power output $P_k^{\text{set}}$ of the HPP is:
\begin{equation}
    {P}^{\text{set}}_{k} = {P}^{\text{disp}}_{k} + \left[ 50 - f_k\right] \cdot \sigma_f
\end{equation}
where the term $f_k$ indicates the grid frequency value at time $k$. As a consequence, the tracking error $\text{TE}_k$ is the difference between the expected power output ${P}^{\text{set}}_{k}$ and $P^{\text{pcc}}_k$:
\begin{equation}
    \text{TE}_{k} = {P}^{\text{set}}_{k} - P^{\text{pcc}}_k
    \label{eq:DispatchTracking_k}
\end{equation}
To assess the effectiveness of the FCR provision, we introduce the \emph{Root Mean Squared} (RMS) of the tracking error array $\mathbf{TE}$, defined in \cref{eq:DispatchTracking}, over the entire window from $k_1$ to $k_2$ :
\begin{equation}
    RMS(\mathbf{TE}_{k_1:k_2}) = RMS\bigg(\mathbf{P}^{\text{set}}_{k_1:k_2} -\mathbf{P}^{\text{pcc}}_{k_1:k_2}\bigg)
    \label{eq:DispatchTracking}
\end{equation}
where $\mathbf{TE}_{k_1:k_2}$ is an array containing the values $\text{TE}_k$ $\forall k \in [k_1,k_2]$. The same nomenclature applies to $\mathbf{P}^{\text{set}}_{k_1:k_2}$ and $\mathbf{P}^{\text{pcc}}_{k_1:k_2}$. 

The RMS of $\mathbf{TE}$, as defined in Equation \cref{eq:DispatchTracking}, is calculated based on the mean value of $TE_k$ for each minute interval. This metric is used to assess the quality of FCR tracking. The use of 1-minute averaging allows for the evaluation of the FCR response, specifically in terms of mean power, within a standard time frame for FCR provision, as specified in \cite{european_commission_commission_2017}.
\subsubsection{Wear and Tear Reduction}
Wear reduction is assessed through three specific KPIs: servomotors mileage, servomotors \emph{ Number of Movements} (NoM) and derivative of \emph{ Runner Blade Torque} (RBT). The reduction in both servomotors milage and NoM serves as a significant indicator of the wear reduction achieved by hybridization \cite{valentin_benefits_2022}. Finally, to estimate the forces affecting the blades and their servomechanisms, the torque on the blade connecting rod is evaluated. In particular, the \emph{Cumulative Density Function} (CDF) of the RBT derivative is analyzed. 
\subsubsection{Hydraulic and global efficiency}
Two key efficiency metrics are considered: global efficiency $\eta^g$ and hydraulic efficiency $\eta^h$. Global efficiency, also known as overall efficiency, encompasses all losses within the energy conversion process, including mechanical, electrical, and hydraulic losses. For the experiments, the hydraulic power for each time $k$ is calculated as a function of the measured discharge $Q_k$ and net head $H_k$:
\begin{equation}
P^h_k = \rho \cdot g \cdot Q_k \cdot H_k
\end{equation}
where $\rho$ indicates the water density (i.e., \SI{1000}{\kilogram\per\cubic\meter}) and $g$ the gravitational acceleration (i.e., \SI{9.81}{\meter\per\second\squared}
). With the knowledge of the electrical power, measured at the PCC, the global efficiency $\eta^g_k$ is estimated as:
\begin{equation}
    \eta^g_k = P^{\text{pcc}}_k / P^h_k
\end{equation}
 On the other hand, the hydraulic efficiency $\eta^h$ specifically quantifies the efficiency of the water-to-mechanical energy conversion within the hydro machine. It is based primarily on the condition of the turbine runner and considers losses associated with hydraulic components.  For the experiments, thanks to measurements of the torque $T_k$ and speed $n_k$ on the shaft, it is possible to compute the mechanical power $P^m_k$ and, therefore, the hydraulic efficiency $\eta^h_k$ as follows:
 \begin{equation}
    P^m_k = T_k \cdot n_k
\end{equation}
\begin{equation}
    \eta^h_k = P^m_k / P^h_k
\end{equation}

\section{Turbine Hill-chart and Control}
The control mechanism for fixed-speed Kaplan turbines relies on the CAM block, which optimizes efficiency by linking the \emph{Runner Blade Angle} (RBA) $\beta$ to the \emph{Guide Vanes Opening} (GVO) $\alpha$ and the head. For Kaplan governors, the CAM block establishes the relationship between GVO and RBA for each head value. In contrast, in variable-speed propellers, when the head remains fixed, the CAM block defines the optimal speed for each GVO, assuming a constant RBA (propellers are not able to change the angle of the blades). To be able to compute both CAM blocks, a model of the turbine efficiency has to be provided.

\subsection{Efficiency Meta-Model}
\label{sec:MetaModels}
The scope of hydraulic turbine surrogate models is to predict machine characteristics, such as hydraulic efficiency $\eta^h$, for every operating point. In particular, since the final goal is the deployment of these models for control purposes, efficiency should be modeled as a function of the controllable and input variables of the system. According to \cite{brezovec_nonlinear_2006}, Kaplan hydraulic $\eta^h$ can be modelled as nonlinear function of:
\begin{enumerate*}[label=(\roman*)]
    \item GVO,
    \item RBA,
    \item the net head $H$, and
    \item the unit rotational speed $n$:
    \end{enumerate*}
\begin{equation}
    \eta^h=\eta^h(\alpha,\beta,H,n)
    \label{eq:eta}
\end{equation}

Starting from (\ref{eq:eta}), the IEC speed coefficient $n_{ED}$ \cite{standard_iec_and_others_iec_2019} is introduced to reduce the number of variables and generalize the problem:
\begin{equation}
    n_{ED} = \frac{n\cdot D}{\sqrt{gH}}
    \label{eq:n11}
\end{equation}
Therefore, the surrogate model of the efficiency $\eta^h$ is built as a function of $\alpha$, $\beta$ and $n_{ED}$:
\begin{equation}
    \eta^h=\eta^h(\alpha,\beta,n_{ED})
    \label{eq:eta2}
\end{equation}
Experimental data is gathered and used to fit \cref{eq:eta2}. Measurements are recorded under stationary conditions, with a constant head of \SI{10}{\meter}, while adjusting the GVO and RBA from zero to full opening in increments of 1 degree. Variable speed tests are conducted within the range of 500 to \SI{1500}{\rpm}, translating into values of $n_{ED}$ between 0.29 and 0.86, as a consequence of \cref{eq:n11}. The recorded data points describe the efficiency of the turbine across the part-load, the best efficiency point, and the full-load operational ranges. The data set is then used to fit the surrogate model $\eta^h$ for the Kaplan runner. For this purpose, similarly to \cite{gerini_optimal_2021}, the {Multivariate Adaptive Regression Spline} technique \cite{friedman_multivariate_1991} is employed. This technique is selected for its ability to fit \cref{eq:eta2} with analytical functions, continuous in their first derivative, which are particularly suitable to be used in optimization frameworks. A similar approach can also be found in \cite{vagnoni_interaction_2019}. The fitting of the model results in a  \emph{Mean Square Error} (MSE) = $0.0009$ and a \emph{Coefficient of Determination} $R^2=0.97$.
\subsection{Kaplan CAM Curve}
\label{sec:CAMcurve}
\begin{figure}[t]
\centering
\includegraphics[width=\linewidth]{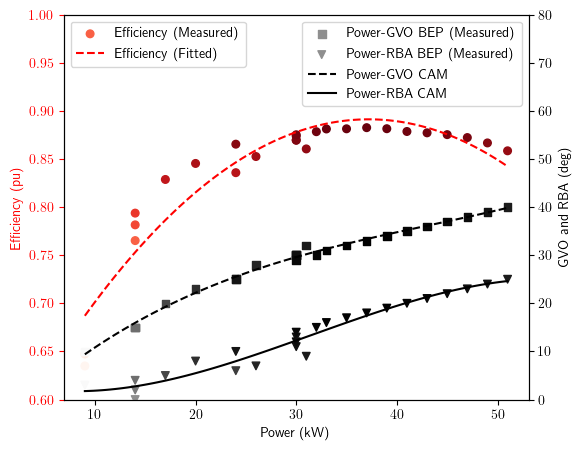}
\caption{CAM Curves and expected efficiency for the Kaplan runner for a constant head $H= \SI{10}{\meter}$ and a fixed speed of $n=\SI{1500}{\min^{-1}}$}
\label{fig:CAM}
\end{figure}

The CAM control block established the optimal relationship between GVO and RBA to maximize efficiency for a given head value and a fixed rotational speed. For each head, the CAM is usually built as a two-column lookup table containing different values of GVO and the corresponding RBA, where each line corresponds to a certain power set-point (for a fixed speed).
Once the efficiency of the Kaplan turbine is expressed as a function of $\alpha$, $\beta$, and $n_{\text{ED}}$ by the meta-model (\ref{eq:eta2}), the CAM is obtained by performing numerical optimization on this function, following the procedure presented in \cite{gerini_optimal_2021}.  The CAM curve for the reduced-scale Kaplan, for a head value of \SI{10}{\meter}, is visible in \cref{fig:CAM}.

\subsection{VARspeed Propeller Governor}
\label{sec:VARspeedGovernor}
\begin{figure}[t]
\centering
\includegraphics[width=\linewidth]{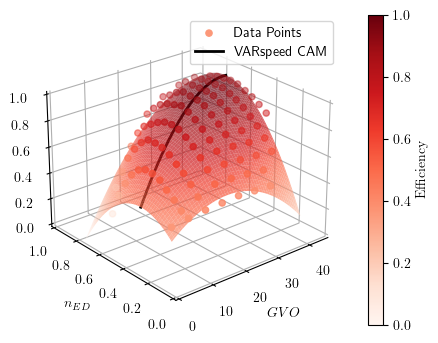}
\caption{Efficiency hill-chart and speed CAM of the VARspeed propeller with RBA = \SI{18}{\degree}}
\label{fig:CAMspeed}
\end{figure}

The Kaplan is retrofitted as a propeller by fixing the RBA to the angle of the blade corresponding to the best efficiency point, i.e., to $\beta=\SI{18}{\degree}$, as visible from \cref{fig:CAM}. For a VARspeed propeller, the governor is responsible for choosing the best speed and GVO for each power set point \cite{borkowski_small_2020,fraile-ardanuy_variable-speed_2006}. Following the same procedure as for the Kaplan CAM curve block, but using as control variable $\alpha$ and $n$, instead of $\alpha$ and $\beta$, the VARspeed CAM can be obtained. The efficiency curves are presented in \cref{fig:CAMspeed}. The figure contains the measured points and a fit of the efficiency meta-model, while the VARspeed CAM is visible in black.

\section{Experimental Results}
\label{sec:Experimental Validation}
\subsection{FCR provision quality}
The results in Table \ref{tab:rmse-values-reduction} provide a comprehensive analysis of the FCR tracking error, computed as shown in \cref{eq:DispatchTracking}, for various control configurations. These configurations include the baseline scenario, i.e. \emph{'Only Hydro'}, together with the BESS hybridization and VARspeed scenarios. The BESS hybrid configuration demonstrates a remarkable reduction when compared to the '\emph{Only Hydro}'. This reduction is particularly evident, with both hybrid systems consistently achieving a reduction of at least 50\% in tracking error compared to the non-hybrid '\emph{Only Hydro}' system. The VAR-speed propeller operation shows slightly worse performance but still manages to reduce tracking error by 14\% compared to the non-hybrid '\emph{Only Hydro}' system.
\begin{table}[ht]
    \centering
    \begin{tabular}{l c}
        \toprule
        \textbf{Configuration} & \textbf{Error (30s)} \\
        \midrule
        % \textcolor{black}{DBF (\SI{5}{\kilo\watt}/\SI{5}{\kilo\watt\hour} BESS)}     & 0.4095 ($-47.48\%$) \\
        % \textcolor{orange}{DLMPC (\SI{9}{\kilo\watt}/\SI{9}{\kilo\watt\hour} BESS)} & 0.3544 ($-54.54\%$) \\
        % \textcolor{green}{DLMPC (\SI{5}{\kilo\watt}/\SI{5}{\kilo\watt\hour} BESS)}   &  0.3696 ($-49.30\%$) \\
        % \textcolor{black}{DLMPC (\SI{9}{\kilo\watt}/\SI{9}{\kilo\watt\hour} BESS)} & 0.3468 ($-60.39\%$) \\
        % \textcolor{purple}{Only Hydro} &  0.7331 ($+\phantom{0}0.00\%$) \\
        % \textcolor{black}{DBF (\SI{5}{\kilo\watt}/\SI{5}{\kilo\watt\hour} BESS)}     & 0.3872 ($-47.18\%$) \\
        % \textcolor{orange}{DBF (\SI{9}{\kilo\watt}/\SI{9}{\kilo\watt\hour} BESS)} & 0.3544 ($-51.65\%$) \\
        \textcolor{green}{DLMPC (\SI{5}{\kilo\watt}/\SI{5}{\kilo\watt\hour} BESS)}   &  0.3696 ($-49.58\%$) \\
        \textcolor{black}{DLMPC (\SI{9}{\kilo\watt}/\SI{9}{\kilo\watt\hour} BESS)} & 0.3468 ($-52.69\%$) \\
        \textcolor{purple}{Only Hydro} &  0.7331 ($+\phantom{0}0.00\%$) \\
        \textcolor{brown}{VARspeed} & 0.6284 ($-14.18\%$) \\
        \bottomrule
    \end{tabular}
    \vspace{0.2cm}
    \caption{RMSE Values and Reduction (\%) of the Tracking Error (TE) with respect to \emph{Only Hydro}.}
\label{tab:rmse-values-reduction}
\end{table}

\subsection{Wear and Tear Reduction}
Table \ref{tab:mileage-movements} compares the wear reduction achieved through the BESS hybridization or VARspeed operation against the one in Kaplan standalone operation. All percentage values are relative to the '\emph{Only Hydro}' baseline configuration. The BESS hybridized Kaplan operation reduces both GVO and RBA mileage and number of movements with respect to the baseline configuration. For example, with a \SI{5}{\kilo\watt} BESS, the hybrid setup achieves a substantial 94.0\% mileage reduction for both GVO and RBA. An even greater reduction is visible for the number of movements (-97\%).  Notably, the VARspeed propeller case exhibits a unique characteristic in which the turbine blades remain stationary, resulting in zero blade movement and mileage for the servomotors (-100\%). Simultaneously, it is essential to acknowledge that VARspeed control significantly increases guide vane movements (+45\%) and mileage (+122\%).
% \begin{table*}[t]
%     \centering
%     \caption{Reduction in Mileage and Movements (\%)}
%     \begin{tabular}{lcccc}
%     \toprule
%     \textbf{Configuration} & \multicolumn{2}{c}{\textbf{Guide Vanes Opening}} & \multicolumn{2}{c}{\textbf{Runner Blade Angle}} \\
%      & \multicolumn{2}{c}{(GVO)} & \multicolumn{2}{c}{(RBA)} \\
%     \cmidrule(lr){2-5}
%       & \textbf{Mileage } & \textbf{Number of Movements} & \textbf{Mileage } & \textbf{Number of Movements} \\
%     \midrule
%     % \textcolor{black}{DBF (\SI{5}{\kilo\watt} BESS)} & 4.18 (-91.5\%) & 662 (-92.9\%) & 3.06 (-90.7\%) & 753 (-91.6\%) \\
%     % \textcolor{orange}{DBF (\SI{9}{\kilo\watt} BESS)} & 0.98 (-98.0\%) & 126 (-98.6\%) & 0.70 (-97.9\%) & 128 (-98.6\%) \\
%     \textcolor{black}{DLMPC (\SI{5}{\kilo\watt}/\SI{5}{\kilo\watt\hour} BESS)} & 3.33 (-93.2\%) & 292 (-96.8\%) & 1.98 (-94.0\%) & 258 (-97.1\%) \\
%     \textcolor{orange}{DLMPC (\SI{9}{\kilo\watt}/\SI{9}{\kilo\watt\hour} BESS)} & 0.99 (-98.0\%) & 64 (-99.3\%) & 0.61 (-98.2\%) & 52 (-99.4\%) \\
%     \textcolor{green}{Only Hydro} & 49.03 (+0.00\%) & 9261 (+0.00\%) & 32.93 (+0.00\%) & 8970 (+0.00\%) \\
%     \textcolor{black}{VARspeed} & 109.09 (+122.50\%) & 13484 (+45.60\%) & 0 (-100.00\%) & 0 (-100.00\%) \\
%     \bottomrule
%     \end{tabular}
%     \label{tab:mileage-movements}
% \end{table*}
\begin{table*}[ht]
    \centering
    \begin{tabular}{lcccc}
    \toprule
    \textbf{Configuration} & \multicolumn{2}{c}{\textbf{Guide Vanes Opening}} & \multicolumn{2}{c}{\textbf{Runner Blade Angle}} \\
     & \multicolumn{2}{c}{(GVO)} & \multicolumn{2}{c}{(RBA)} \\
    \cmidrule(lr){2-5}
      & \textbf{Mileage } & \textbf{Number of Movements} & \textbf{Mileage } & \textbf{Number of Movements} \\
    \midrule
    % \textcolor{black}{DBF (\SI{5}{\kilo\watt} BESS)} & 4.18 (-91.5\%) & 662 (-92.9\%) & 3.06 (-90.7\%) & 753 (-91.6\%) \\
    % \textcolor{orange}{DBF (\SI{9}{\kilo\watt} BESS)} & 0.98 (-98.0\%) & 126 (-98.6\%) & 0.70 (-97.9\%) & 128 (-98.6\%) \\
    \textcolor{green}{DLMPC (\SI{5}{\kilo\watt} BESS)} & 3.33 (-93.2\%) & 292 (-96.8\%) & 1.98 (-94.0\%) & 258 (-97.1\%) \\
    \textcolor{black}{DLMPC (\SI{9}{\kilo\watt} BESS)} & 0.99 (-98.0\%) & 64 (-99.3\%) & 0.61 (-98.2\%) & 52 (-99.4\%) \\
    \textcolor{purple}{Only Hydro} & 49.03 (+0.0\%) & 9261 (+0.0\%) & 32.93 (+0.0\%) & 8970 (+0.0\%) \\
    \textcolor{brown}{VARspeed} & 109.09 (+122.5\%) & 13484 (+45.6\%) & 0.0 (-100.0\%) & 0.0 (-100.0\%) \\
    \bottomrule
    \end{tabular}
    \vspace{0.2cm}
    \caption{Reduction in Mileage and Movements (\%) of the GVO and RBA servomechanisms.} 
    \label{tab:mileage-movements}  % Add the label here
\end{table*}
% Finally, for an estimate of the forces that affect the blades and their servomechanisms, the derivative of the torque occurring on the runner blades' servomechanism is evaluated (see \cref{fig:RBT}). In line with the trends observed in other wear reduction KPIs, the \emph{'Only Hydro'} case demonstrates inferior performance, characterized by elevated levels of torque derivative on the blades. In the case of the 9 kW BESS, it is noticeable that the CDF of the RBT derivative is significantly narrower. This narrower distribution implies that there are fewer occurrences of high torque derivative, indicating improved performance in mitigating strong torque changes with BESS hybridization. The presence of a larger BESS size demonstrates a positive impact on the reduction of torque derivative. The CDF of the RBT derivative is even narrower for the VARspeed case, outperforming both BESS hybridization cases and minimizing these torque oscillations.

\begin{figure}[ht]
    \centering
    \includegraphics[width=\linewidth]{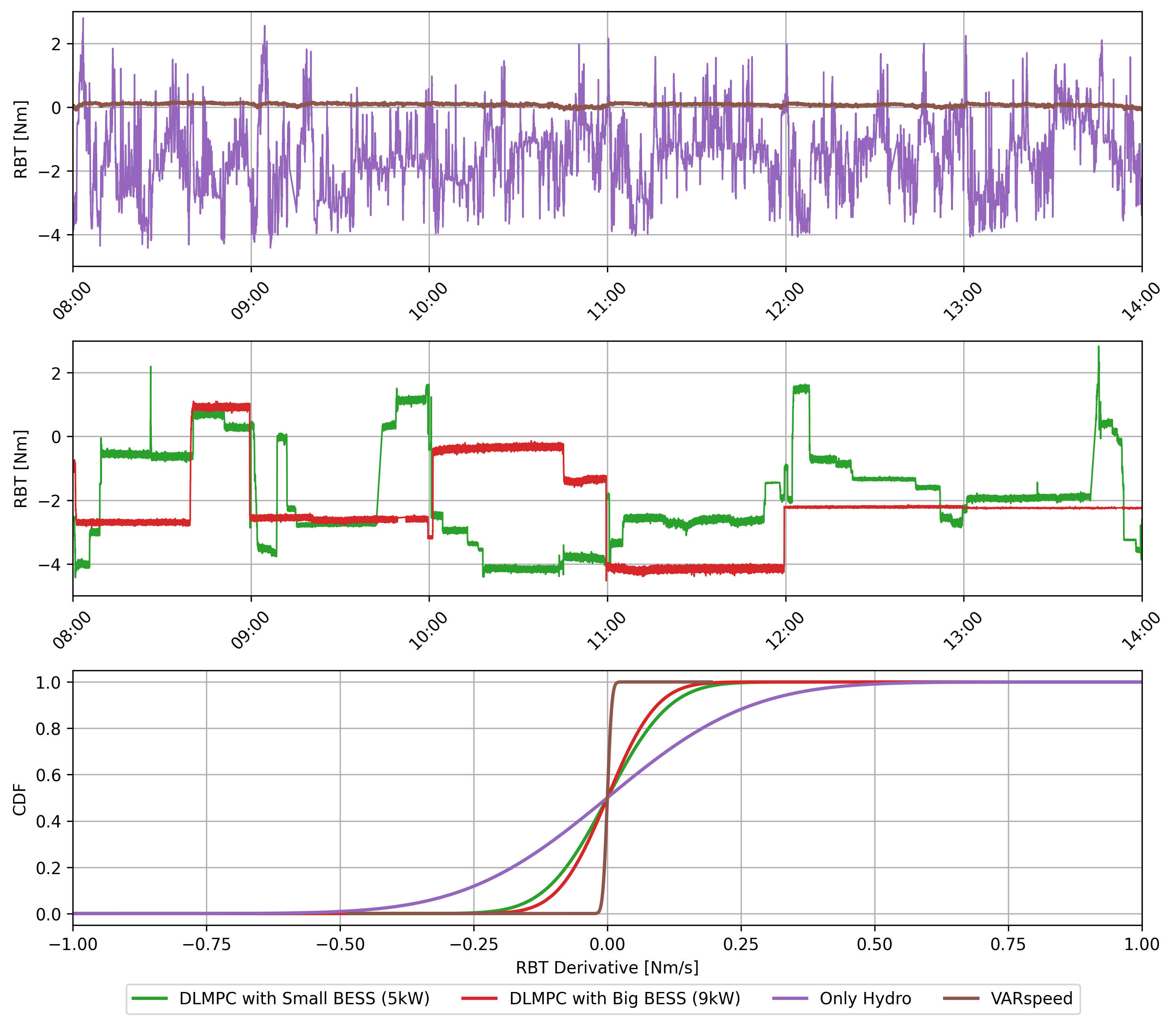}
    \caption{\textcolor{black}{Comparative analysis of blade torque oscillation across all test scenarios.}}
    \label{fig:RBT2}
\end{figure}

% \begin{figure}[t]
%     \centering
%     \includegraphics[width=\linewidth]{pscc2024_template_LaTex-2/Figures/VARspeed_RBT.png}
%     \caption{Comparative analysis of RBT derivative all test scenarios.}
%     \label{fig:RBT}
% \end{figure}
% % \begin{figure}[t]
% %     \centering
% %     \includegraphics[width=0.9\linewidth]{pscc2024_template_LaTex-2/Figures/VARspeed_ShaftTorqueDerivative.png}
% %     \caption{Comparative analysis of Shaft Torque derivative all test scenarios.}
% %     \label{fig:ShaftTorque}
% % \end{figure}
Finally, for an estimate of the forces that affect the blades and their servomechanisms, the derivative of the torque occurring on the runner blades' servomechanism is evaluated. \textcolor{black}{\cref{fig:RBT2} shows the blade torque measurement for the first \SI{6}{\hour} of the experiment in the upper subplots and the CDF of the RBT derivative on the lower one. In line with the trends observed in other wear reduction KPIs, the \emph{'Only Hydro'} case demonstrates inferior performance, characterized by elevated levels of torque derivative on the blades.} The CDF of the RBT derivative is most constricted in the VARspeed case, surpassing both BESS hybrid setups in reducing torque variations. Specifically, \cref{fig:RBT2} highlights the significant advantage of lacking the RBA servomechanism. Indeed, for the VARspeed scenario, the torque on the blade pin maintains a near-constant level (as visible from the first subplot), signifying minimal torque fluctuations throughout the operation. \textcolor{black}{It is important to note that the mechanism operating the runner blades in the reduced-scale experimental setup does not perfectly mimic the prototype's mechanism, primarily due to spatial constraints within the reduced-scale model that limit the installation of an identical servo-mechanism. As a result, within the context of this study, the RBT measurements are utilized primarily as a qualitative indicator rather than a quantitative KPI that can be directly scaled to prototype dimensions.}
\subsection{Hydraulic and Global Efficiency}
\label{sec:ExperimentalResults:Efficiency}
\begin{figure*}[ht]
    \centering
    \begin{minipage}{0.4\textwidth}
        \centering
        \includegraphics[width=\linewidth]{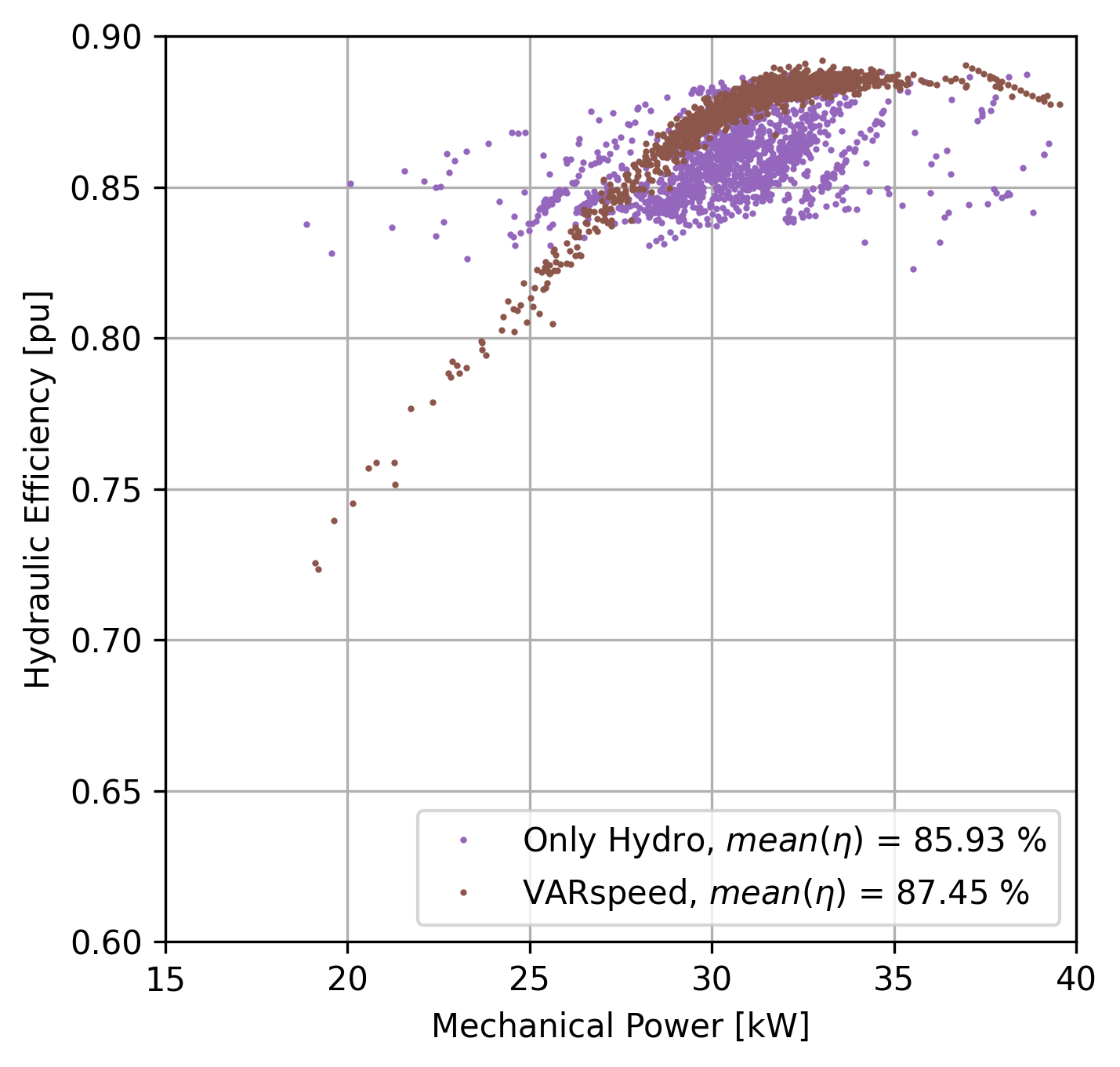}
        \caption{Hydraulic Efficiency $\eta$.}
        \label{fig:HydraulicEfficiency}
    \end{minipage}%
    \begin{minipage}{0.4\textwidth}
        \centering
        \includegraphics[width=\linewidth]{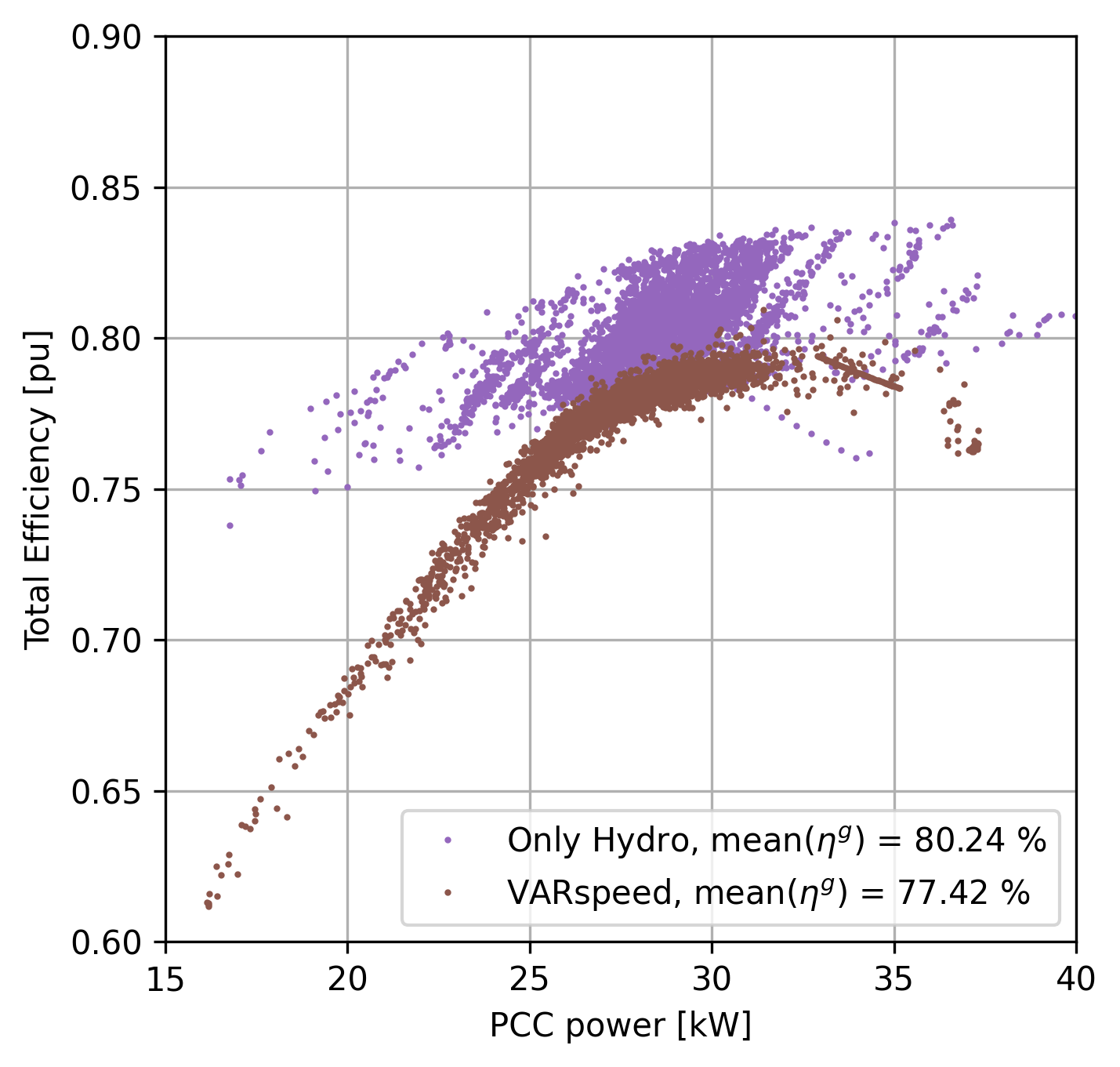}
        \caption{Global Efficiency $\eta^g$.}
        \label{fig:TotalEfficiency}
    \end{minipage}
\end{figure*}

% \cref{fig:HydraulicEfficiency} illustrates the hydraulic efficiency of the fixed-speed Kaplan turbine and the VARspeed propeller (fixed blades) across the range of operating conditions tested. The graph demonstrates that the VARspeed propeller exhibits slightly higher hydraulic efficiency on average (+ 1.5\%), despite being less efficient in part-load conditions. This is caused by the choice to fix the Kaplan RBA to \SI{18}{\deg}, close to the power of the best efficiency point, as visible from \cref{fig:CAM}. Overall, the increase in hydraulic efficiency by the VARspeed control strongly depends on the point of operation during the test. Hence, we cannot definitively claim that the VARspeed operation of a retrofitted Kaplan turbine as a propeller is superior in efficiency to its original Kaplan operation. However, we can at least consider them to be viable alternatives. The hydraulic efficiency of the fixed-speed Kaplan, shown in \cref{fig:HydraulicEfficiency} in green, is in line with the expected efficiency values of \cref{fig:CAM}. It is worth noticing that these efficiencies can be considered representative of the prototype machine in hydraulic similarity to the reduced-scale model tested.
\cref{fig:HydraulicEfficiency} illustrates the hydraulic efficiency of the fixed-speed Kaplan turbine and the VARspeed propeller (fixed blades) across the range of operating conditions tested\footnote{\textcolor{black}{\cref{fig:HydraulicEfficiency} shows multiple efficiency values for each power value because of head variation related to the dynamic of the testing. This is a clear limitation when drawing conclusions on efficiency for dynamic tests.}}. The graph demonstrates that the VARspeed propeller exhibits slightly higher hydraulic efficiency on average (+ 1.5\%), despite being less efficient in part-load conditions. This is caused by the choice to fix the Kaplan RBA to \SI{18}{\degree}, close to the power of the best efficiency point, as visible from \cref{fig:CAM}. Overall, the increase in hydraulic efficiency by the VARspeed control strongly depends on the operating point. Hence, we cannot definitively claim that the VARspeed operation of a retrofitted Kaplan turbine as a propeller is superior in efficiency to its original Kaplan operation. However, we can at least consider them to be viable alternatives. The hydraulic efficiency of the fixed-speed Kaplan, shown in \cref{fig:HydraulicEfficiency} in green, is in line with the expected efficiency values of \cref{fig:CAM}. \textcolor{black}{It is worth noticing that these efficiencies can be considered representative of the prototype machine in hydraulic similarity to the reduced-scale model tested since the tests are performed in compliance with the IEC 60193 standards \cite{standard_iec_and_others_iec_2019} efficiency measurements in reduced scale model testing.}
% \begin{figure}[t]
% \centering
% \includegraphics[width=\linewidth]{pscc2024_template_LaTex-2/Figures/VARspeed_HydraulicEfficiency.png}
% \caption{Hydraulic Efficiency $\eta^h$ of the \emph{Only Hydro} setup compared to VARspeed}
% \label{fig:HydraulicEfficiency}
% \end{figure}
% \begin{figure}[t]
% \centering
% \includegraphics[width=\linewidth]{pscc2024_template_LaTex-2/Figures/VARspeed_TotalEfficiency.png}
% \caption{Global Efficiency $\eta^g$ of the \emph{Only Hydro} compared to VARspeed}
% \label{fig:TotalEfficiency}
% \end{figure}

In \cref{fig:TotalEfficiency}, the global efficiency for the baseline case and the VARspeed propeller are displayed. The global efficiency includes the efficiency of the generator for the fixed-speed case and the combined efficiency of the generator and full-size frequency converter for the variable-speed case. The graph shows that, when accounting for the losses incurred by the power converter, the global efficiency of variable-speed propeller turbines tends to be lower compared to the fixed-speed Kaplan. However, the efficiency of the synchronous machine and the converter do not scale to the prototype, making it hard to conduct a precise quantitative analysis.
\newline

In conclusion, \cref{tab:final_comparison} summarizes the performance of each configuration. In particular, the "+" or "-" symbols indicate whether a configuration performs better or worse, respectively, than the baseline \emph{"Only Hydro"} scenario. When both configurations outperform the baseline, we use "+\,+" to highlight the superior performer. In the "FCR" column, the symbol "+" signifies more accurate tracking of the FCR power, indicating a smaller error. Regarding wear and tear, the "+" value denotes a greater reduction in wear, resulting in less mileage, reduced movements, and lower values of the RBT derivative. In the latter columns, the symbol "+\,-" is used to indicate similar performance to the baseline configuration.
\begin{table}[ht]
    \centering
  \begin{threeparttable}
    \caption{KPIs Comparison with the 'Only Hydro' Scenario}
    \begin{tabular}{l | c | c c c | c c}
    \toprule
    \textbf{Configuration} & {\textbf{FCR}} & \multicolumn{3}{c}{\textbf{Wear and Tear}} & \multicolumn{2}{c}{\textbf{Efficiency}} \\
    & $\text{RMS}(\mathbf{TE})$ & GVO & RBA & RBT &  $\eta^h$ & $\eta^g$\\
    \midrule
    BESS Hybrid HPP  & +\,+ & + & + &  + & +\,-  & +\,- \\
    % Only Hydro & 0 & 0 & 0 & 0 & 0 & 0 & 0 \\
    VARspeed & + & - & +\,+ & +\,+  & +\footnotemark[1]   & -\footnotemark[2]\\
    \bottomrule
    \end{tabular}
\label{tab:final_comparison}
     \begin{tablenotes}
      \item[1] As discussed in \cref{sec:ExperimentalResults:Efficiency}, this result strongly depends on the operating points of the hydraulic turbine. It is not possible to claim an overall increase of $\eta^h$ when converting a Kaplan turbine into a VARspeed propeller. 
      \item[2] It is not possible to scale $\eta^g$ to the prototype. However, it is clear that the presence of the full-size frequency converter results in a decrease in the overall efficiency.
    \end{tablenotes}
  \end{threeparttable}
\end{table}

\section{Conclusion}
\label{sec:Conclusion}
In this paper, we discuss and compare different solutions to address \emph{Run-of-River Hydropower Plants} (RoR HPPs) providing \emph{Frequency Containment Reserve} (FCR), while trying to reduce mechanical stresses and prolong the operational life of its components. Two approaches are considered and compared to the baseline scenario ('Only Hydro'): BESS hybridization of the Kaplan unit, with a dedicated control framework to split the power set-point between BESS and HPP, and variable speed control of the Kaplan operated as a propeller, i.e., with fixed blades.
Comprehensive reduced-scale experiments have been conducted in an innovative testing platform for the two approaches, under the same grid scenario. The hybridized system showcased enhanced FCR provision quality and a notable reduction in servomechanism stress. The analysis of wear reduction revealed a substantial decrease in both guide vane and runner blade mileage, ranging from 93.2\% with a 5 kW BESS to up to 100\% for the \emph{Variable Speed} (VARspeed) propeller. Similar reductions were observed in the case of the number of movements. Moreover, the reduced torque derivative on the blades further emphasizes the benefits of both approaches. Considering the sensitivity of the Kaplan turbine's blade mechanism, a conclusive preference between the two setups is not evident. The choice depends largely on the specific conditions of the power plant. When the Kaplan turbine is nearing the end of its operational life and the blade servomotors are no longer functional, VARspeed control offers an attractive solution that does not require any blade movement. In contrast, when the Kaplan turbine is aging but still operational, the BESS hybrid configuration stands out, extending the turbine's life while ensuring efficient power generation. This emphasizes the importance of a case-dependent approach, aligning CAPEX investment strategies with the machine's conditions. \textcolor{black}{Future work will include a detailed economic feasibility study, examining potential costs and benefits associated with both BESS hybridization and VARspeed conversion of aging HPP assets.}
\ifCLASSOPTIONcaptionsoff
  \newpage
\fi

% \section{Conference paper preparation}
% Papers accepted for PSCC are normally no longer than 7 pages. However, in exceptional cases, up to 2-3 additional pages can be accepted if the extra length is required to fully but still succinctly develop and explain an idea and its application. In such cases, the authors will be required to briefly explain why the extra length is necessary. 

\bibliographystyle{IEEEtran}

\bibliography{references.bib}

% \appendices
\end{document}